\documentclass[twocolumn,backref, breaklinks, colorlinks]{aastex63}
\hypersetup{linkcolor=blue,citecolor=blue,filecolor=cyan,urlcolor=magenta}
\usepackage{tabularx}
\usepackage{url}
\usepackage{graphicx}
\usepackage[T1]{fontenc}
\usepackage{ae,aecompl}
\usepackage{booktabs}
\usepackage{natbib}
\usepackage{longtable}

\usepackage{blindtext}
\usepackage{longtable}
\usepackage{tabu}
\usepackage{lineno}

\def\unit #1{\,{\rm #1}}

\newcommand\kev{\rm \,\unit{keV}}

\newcommand\funit{\rm \,erg\,cm^{-2}\,s^{-1}}

\newcommand\ba{{\it BatAnalysis} }

\newcommand\chandra{{\it Chandra}}

\newcommand\xmm{{\it XMM-Newton}}

\newcommand\suzaku{{\it Suzaku}}
\newcommand\nustar{{\it NuSTAR}}

\def\aap{A\&A} \def\apjl{ApJ}  
 \def\apjs{ApJS}

\submitjournal{ApJ}

\begin{document}

	\title{ BatAnalysis - A Comprehensive Python Pipeline for Swift BAT Survey Analysis}

	\author[0000-0002-4299-2517]{Tyler Parsotan}
	\affiliation{Center for Space Science and Technology, University of Maryland Baltimore County, 1000 Hilltop Circle, Baltimore, MD 21250, USA.}
	\affiliation{Astrophysics Science Division, NASA Goddard Space Flight Center,Greenbelt, MD 20771, USA.}
	\affiliation{Center for Research and Exploration in Space Science and Technology, NASA/GSFC, Greenbelt, Maryland 20771, USA}

	\author[0000-0003-2714-0487]{Sibasish Laha}
	
	\affiliation{Center for Space Science and Technology, University of Maryland Baltimore County, 1000 Hilltop Circle, Baltimore, MD 21250, USA.}
	\affiliation{Astrophysics Science Division, NASA Goddard Space Flight Center,Greenbelt, MD 20771, USA.}
	\affiliation{Center for Research and Exploration in Space Science and Technology, NASA/GSFC, Greenbelt, Maryland 20771, USA}
	
	\author[0000-0001-7128-0802]{David M. Palmer}
	\affiliation{Los Alamos National Laboratory, Los Alamos, NM 87544, USA}
	
	\author[0000-0002-7851-9756]{Amy Lien}
	\affiliation{University of Tampa, Department of Chemistry, Biochemistry, and Physics, 401 W. Kennedy Blvd, Tampa, FL 33606, USA}

	\author[0000-0003-1673-970X]{S. Bradley Cenko}
	\affiliation{Astrophysics Science Division, NASA Goddard Space Flight Center,Greenbelt, MD 20771, USA.}
	\affiliation{Joint Space-Science Institute, University of Maryland, College Park, MD 20742, USA}
	
	\author[0000-0003-4348-6058]{Hans Krimm}
	\affiliation{National Science Foundation 2415 Eisenhower Ave., Alexandria, VA 22314}

	\author[0000-0001-9803-3879]{Craig Markwardt}
	\affiliation{Astrophysics Science Division, NASA Goddard Space Flight Center,Greenbelt, MD 20771, USA.}
	
	\correspondingauthor{Tyler Parsotan}
	\email{parsotat@umbc.edu}
	

	\begin{abstract}
		
		The Swift Burst Alert Telescope (BAT) is a coded aperture gamma-ray instrument with a large field of view that primarily operates in survey mode when it is not triggering on transient events. The survey data consists of eighty-channel detector plane histograms that accumulate photon counts over time periods of at least 5 minutes. These histograms are processed on the ground and are used to produce the survey dataset between $14$ and $195$ keV. Survey data comprises $> 90\%$ of all BAT data by volume and allows for the tracking of long term light curves and spectral properties of cataloged and uncataloged hard X-ray sources. Until now, the survey dataset has not been used to its full potential due to the complexity associated with its analysis and the lack of easily usable pipelines.
		Here, we introduce the \ba python package\added{, a wrapper for HEASoftpy, } which provides a modern, open-source pipeline to process and analyze BAT survey data. \ba allows members of the community to use BAT survey data in more advanced analyses of astrophysical sources including pulsars, pulsar wind nebula, active galactic nuclei, and other known/unknown transient events that may be detected in the hard X-ray band. We outline the steps taken by the python code and exemplify its usefulness and accuracy by analyzing survey data from the Crab Nebula,  NGC 2992, and a previously uncataloged MAXI Transient. The \ba package allows for $\sim$ 18 years of BAT survey to be used in a systematic way to study a large variety of astrophysical sources.  \\
		
	\end{abstract}
	
	\keywords{}
	
	

	\section{Introduction}\label{sec:intro}
	The Neil Gehrels Swift Observatory \citep{gehrels2004swift} was launched on November 20th, 2004 with the X-ray Telescope (XRT; \citep{burrows_XRT}), the Ultraviolet-Optical Telescope (UVOT; \citep{roming_UVOT}), and the Burst Alert Telescope (BAT; \cite{barthelmy_BAT}) on board. The Swift BAT was fine tuned to detect and localize Gamma-Ray Bursts (GRBs) and has significantly advanced our understanding of these transient events. The BAT uses the coded mask technique to produce small localization regions ($\sim 3$ arc-minutes) and accurate background estimations while maintaining a large field of view ($\sim60^\circ \times 120^\circ$). When a GRB is triggered onboard by BAT, the spacecraft autonomously slews such that the other telescopes can observe the region of the sky where the GRB was localized to. Additionally, the associated BAT data is downlinked to the ground. This data, known commonly as event data, is the highest quality data collected by BAT but is not able to be continuously downlinked. When GRBs are not triggering BAT, the telescope operates in survey mode. This mode compresses the event data that has been collected over some time period (typically 300 sec) into histograms comprising eighty energy channels for each detector element. This data, known as detector plane histograms (DPHs), \replaced{can be}{is} stored \added{onboard} and \added{gets} sent to the ground on a regular basis. 
	
	The BAT survey data  has been used to study X-ray sources in high latitude regions of the galactic plane \cite{markwardt_high_lattitude}, Active Galactic Nuclei \citep{tueller_agn_survey}, and even place limits on transient phenomena \citep{Laha_2022_FRB}. Periodically, the catalog of available survey data has been analyzed in a systematic fashion. These are the 22 month \citep{tueller_22_mo_survey}, 70 month \citep{baumgartner_70_mo_survey}, 105 month \citep{ohr_105_mo_survey}, and 157 month (Lien et. al. in Prep)
	analyses where mosaiced source light curves and spectra, in the energy range of $14-195$ keV, have been published on public websites\footnote{http://swift.gsfc.nasa.gov/results/bs22mon/ \\http://swift.gsfc.nasa.gov/results/bs70mon/ \\ http://swift.gsfc.nasa.gov/results/bs105mon/ }.  The mosaic light curves and spectra from these analyses are typically binned into one month time bins which limits its usefulness to the general astrophysics community. 
	
	\replaced{The method to analyze BAT survey data has been made public through the HEASoft \texttt{batsurvey} pipeline script, however use of this script is not intuitive and has hindered the use of this data by the astrophysics community. Furthermore, the more advanced mosaicing analyses, where the survey data are ``time-integrated'', have not been released publicly due to the complexity associated with this analysis method and the large computational resources that have been needed in the past for producing mosaiced results.}{		While the method to analyze BAT survey data had been made public through the HEASoft \texttt{batsurvey} pipeline script, the more advanced mosaicing analysis, where the survey data are ``time-integrated'', has not been released publicly due to the complexity associated with this analysis method and the large computational resources that have been needed in the past for producing mosaiced results.} With the current computational power available on personal computers and the rise of python packages, we have developed a python package to facilitate the analyses of BAT survey data including producing mosaiced results.  
	
	The \ba package is open source and provides a python interface to process and analyze BAT survey data. This software allows users to use survey data to analyze the long term evolution of known and recently discovered astrophysical sources and, where appropriate, place upper limits on these sources as well. These analyses can be done on a number of different time binnings, from the intrinsic 300 second time binning offered by the DPHs to longer user-defined time binnings that is used to produce mosaic light curves and spectra. 
	
	The structure of this paper is as follows. Section  \ref{survey_heasoft} outlines the general analyses of BAT survey data including the mosaicing process. Section \ref{ba_code} discusses the \ba python package and how it can be used. Section \ref{Sec:discussion} presents the results of our \ba code as compared to prior analyses and shows new analyses that are possible with the software. Finally, in Section \ref{Sec:conclusions}, we conclude with future improvements of the \ba code. 
	
	

	
	
	

	\section{BAT Survey Data} \label{survey_heasoft}
	Here, we describe the general methodology of how the survey data is processed and how mosaic images are calculated. 
	
	\subsection{Processing of Survey Data} \label{survey_orig}
	The pipeline for processing BAT Survey data is implemented in the HEASoft \texttt{batsurvey} script. Here, we briefly describe the steps followed by this script \footnote{Additional information can also be found at: https://swift.gsfc.nasa.gov/analysis/bat\_swguide\_v6\_3.pdf and https://swift.gsfc.nasa.gov/analysis/BAT\_GSW\_Manual\_v2.pdf}. 
	
	Since the survey data come from DPHs that have been integrated for at least 300 seconds, we need to be able to convert these DPHs to detector plane images (DPIs) from which we can extract spectra and fluxes. Before the histograms are converted to sky images, they need to be energy corrected and have spatial and temporal filters applied. 
	
	Starting with the downloaded survey data, the DPHs are adjusted such that all the detectors use the same energy scale with the HEASoft \texttt{batsurvey-erebin} script. After, the DPHs are rebinned into the user-requested energy ranges. By default, the energy ranges are: 14-20, 20-24, 24-35, 35-50, 50-75, 75-100, 100-150, and 150-195 keV. 
	
	The DPHs are then filtered to only include data that occurs during good time intervals (GTIs) and data that is obtained by detectors that produce quality data for the determined GTIs. The GTI filtering is applied with the  \texttt{batsurvey-gti} HEASoft task. The conditions for a time period being a GTI are:
	\begin{itemize}
		\item Swift must be in a stable pointing (the control attitude ``10 arcmin settled'' flag must be set)
		\item The star tracker must be reporting ``OK''
		\item The BAT boresight must be $>30^\circ$ above the Earth's limb
		\item The overall event rate of the detector array is not too high or low, which can be specified by the user with the \texttt{rateminthresh} and \texttt{ratemaxthresh} parameters
		\item a minimum number of detectors are enabled, which can be specified by the user with the \texttt{detthresh}  parameter
		\item no DPH bins have any missing data reported 
		\item the DPH time interval does not cross the midnight UTC boundary\footnote{This restriction is enforced to reject data from this time period where the spacecraft has been commanded to perform small maneuvers in the past. Currently, observations typically end by  00:00:00 UT so this criteria is meant mostly for historical data.}
		\item the spacecraft pointing does not change by more than 1.5 arcmin in pointing and 5 arcmin in roll during each DPH's time interval 
		\item the minimum DPH time interval is 300 seconds
	\end{itemize}
	
	The finest time resolution of a DPH is 300 s, however the integration time for a survey image can last longer.  Processing these different time integrations of DPHs can be specified with the \texttt{timesep} parameter which tells \texttt{batsurvey} to process the data at the 300 second ``DPH'' timescale or the longer ``snapshot'' timescale. Here, we generally refer to the different time resolutions as snapshots. There can be multiple snapshots within a given survey observation ID. 
	
	With the aforementioned filters applied, our DPHs become DPIs. We now need to ensure that the data does not come from bad quality detectors, which uses the HEASoft \texttt{batsurvey-detmask} task. These bad quality detectors include those that:
	\begin{itemize}
		\item have been turned off by the flight software,
		\item are identified as ``hot'' by using the \texttt{bathotpix} algorithm to search the DPI,
		\item have known noisy properties, such as high variance compared to Poisson statistics.
	\end{itemize}
	The detectors that meet these conditions are masked and are not included in the analysis of the DPI. 
	
	At this stage, it is possible to subtract a fixed pattern noise map from the DPIs. This pattern map noise needs to be calculated following the methods outlined in \cite{tueller_22_mo_survey, baumgartner_70_mo_survey, ohr_105_mo_survey}.  As outlined in  \cite{tueller_22_mo_survey}, the pattern noise is spatial and temporal noise that originates from non-uniform detector properties. This type of noise needs to be obtained through lengthy analyses of the detector plane in time and, as a result, is not handled by the \texttt{batclean} script. This source of noise in the DPIs is relevant on the timescales of $\gtrsim$days and is especially important for the creation of the mosaiced images (see section \ref{mosaic_orig}).
	
	With both temporal and spatial filters applied, the DPIs are cleaned. This cleaning is done using \texttt{batclean} and it tries to calculate the contribution of bright sources (with detection SNR greater than the \texttt{cleansnr} parameter) and the background to the counts in the image. This is important since known bright sources can contribute to the detected counts of other sources in the image. This contribution is determined by ray tracing calculations of the shadow pattern of the bright sources on the detector plane. The DPIs are also ``balanced'' at this stage which entails removing the systematic count rate offsets between different regions of the detector plane. The variation from detector to detector are attributed to the variation in the quality of the CdZnTe detectors and the variation in dead times. The variance between detectors located in the inner region of the detector array versus the outer edge of the array is attributed to cosmic ray events and X-ray illumination of the sides of the detectors. The bright sources will also cause the mask support structure to cast a shadow on the detector plane, which is highly energy dependent. To ignore these shadows, we ignore the mask edge regions of the detector for bright sources as determined from  ray trace calculations for bright sources defined by the \texttt{brightthresh} parameter. 
	
	After all this filtering, \texttt{batsurvey} subtracts the bright sources and the background from the DPI and determines if there are any detectors that have counts that deviate from the mean by $N\sigma$, where N is specified by the \texttt{badpixthresh} parameter. The number of remaining detectors then needs to be compared to the user defined \texttt{detthresh2} parameter to determine if the DPI should be kept or not.
	
	Once DPIs have been filtered, we can construct the corresponding sky image. This is done with the HEASoft \texttt{batfftimage} task which cross correlates the DPI with the coded mask pattern. The sky images cover a solid angle of $\sim120^\circ \times \sim 60^\circ$ and are corrected for geometric projection and partial coding effects. The partial coding maps and noise maps are also generated, where the partial coding map denotes the partial exposure of each pixel in the sky image and the noise maps show the root mean square error of the pixel values in the sky image. 
	
	The output image has a resolution of 8.6 arcmin/pixel on-axis and becomes finer by a factor of $\sim 2$ at the edges of the sky image (and associated maps) due to the fact that the sky projection is done in the tangent plane. 
	
	Each sky image can also be analyzed with the HEASoft \texttt{batcelldetect} task, which uses a sliding annulus to search for sources in the image. This task will take in an input catalog consisting of sources and their RA/Dec coordinates and search the image for them. It can also be used to search the image for sources not specified in the input catalog at a SNR above a user-specified value. The output of this task running as a part of \texttt{batsurvey} is a fits catalog with the detected sources and some properties including rates, SNR, and other standard columns produced by the \texttt{batcelldetect} task.
	
	The \texttt{batsurvey} script does not include the capability to calculate a source's spectrum. To calculate the spectrum we first extract the rates for the source of interest in each energy bin and save it to a pulse height amplitude (PHA) file. This information is found in the catalog file output by the \texttt{batsurvey} task. It is important to note that these rates are already background subtracted.  Then, the user needs to calculate the detector response matrix (DRM) for the PHA file corresponding to the snapshot which is done with the \texttt{batdrmgen} script\footnote{BAT measures photons and converts them to pulse heights. A calibration pulser is then used to convert the pulse height back to photon energy. The \texttt{batdrmgen} script attempts to simulate the pulse height that would be expected from a mono-energetic photon beam incident on the detector at the midpoint of each energy bin used in the spectra.}. With these steps complete, the PHA file can be loaded in Xspec and further analysis can be done \citep{xspec}. 
	
	Combining the rates from each snapshot's catalog can provide an energy dependent count rate light curve for our source of interest at either the DPI or snapshot time resolution. Analyzing each spectra allows us to understand any spectral/flux evolution of the source at the same time resolution as well. 
	
\begin{figure*}[th!]
		\centering
		\includegraphics[width=\textwidth]{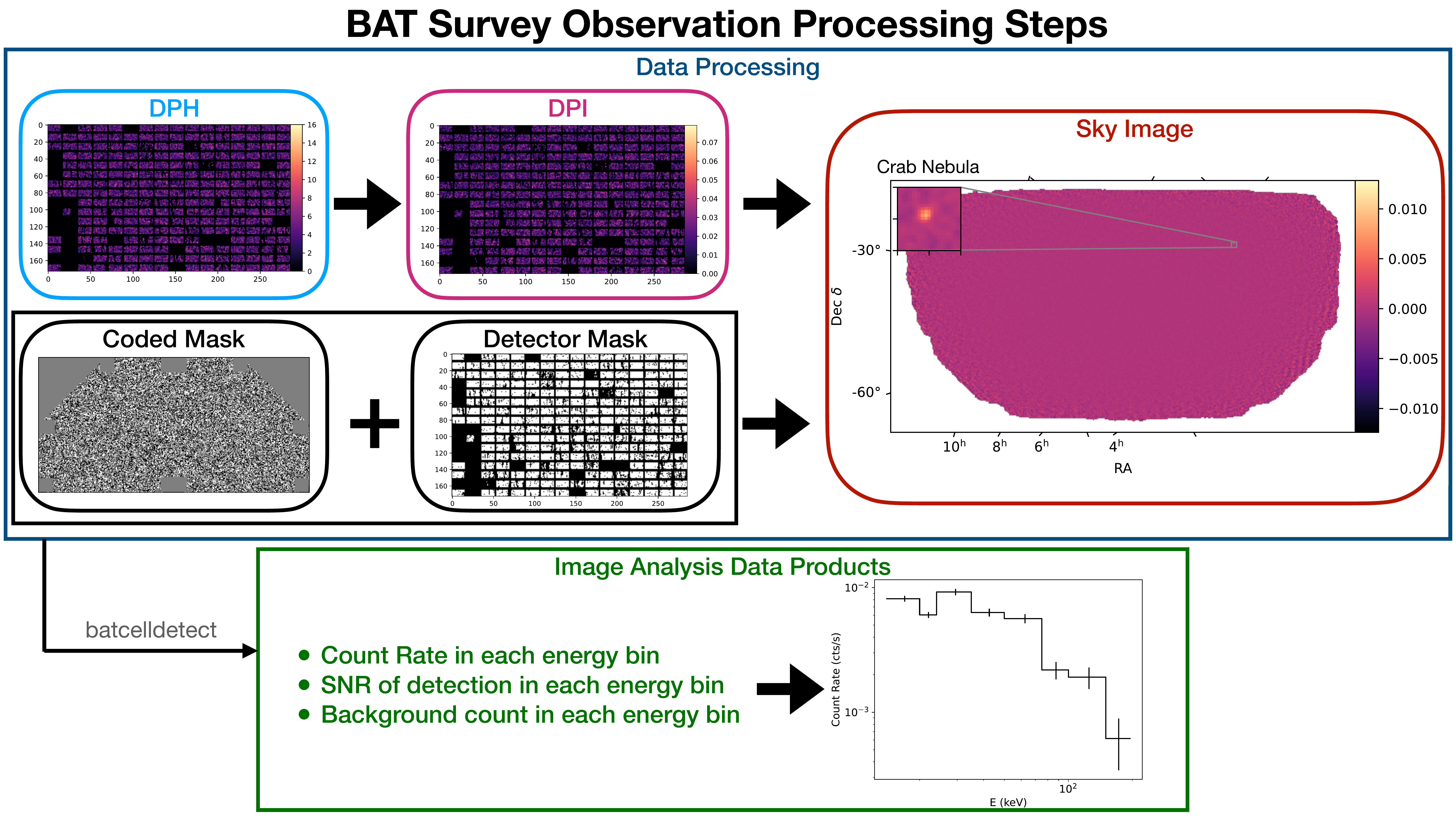}
		\caption{Here, we show an example analysis of a BAT survey dataset that contains the Crab Nebula within the BAT field of view. The survey data starts as a detector plane histogram (DPH) \added{which has units of counts for each detector in the plane}. Then good time interval (GTI) filtering and energy corrections are made, leading to the energy dependent detector plane image (DPI) \added{which has units of count rate for each detector}.  The DPI is \added{further} processed \deleted{a bit more} to remove additional sources of noise and then it is convolved with the BAT coded mask and the detector mask to produce the sky image with units of count rate in each of the 8 energy bands. Then, using \texttt{batcelldetect}, the sky image is processed and sources are identified. The source count rate in each energy band is calculated as well as the background count rate allowing for spectra to be produced,  energy dependent light curves, and the SNR of the detection in each energy band.  }
		\label{fig:flowchart}
	\end{figure*}

	The steps outlined here are illustrated in Figure \ref{fig:flowchart} where we show an example of processing BAT survey data for a single observation ID of the Crab Nebula.  \added{The \ba package follows these same steps to process BAT survey data, as is outlined in Section \ref{ba_code}.}
	
	\subsection{Mosaiced Images} \label{mosaic_orig}
	The processing presented in the previous section represents a ``time-resolved`` survey analysis. Various survey observations can be combined to conduct ``time-integrated`` survey analyses but the individual survey data needs to be combined with systematic noise taken onto account (for example with noise pattern maps). In the case of noise pattern maps, these are time-dependent maps of noise in the detector plane that needs to be subtracted otherwise the noise pattern will accumulate as images are added together. Thus, the noise pattern maps need to be passed into \texttt{batsurvey} before doing a mosaicing analysis where the passed in pattern map is one that has been calculated closest in time to the survey observation that is being analyzed. Additionally the combination of survey observations must be done in a specific way to minimize statistical errors and maximize the SNR of faint sources. 
	
	To calculate our mosaic image, we first need to bin our survey observations into time bins of interest and then add all the snapshots together to get the final exposure map, the partial coding map, the sky image, and the variance map. The individual snapshots are linearly interpolated onto grids of the entire sky that uses the zenith equal angle projection to prevent distortion of BAT's point spread function. There are 6 sky facets which overlap with one another and each has a central pixel resolution of $\sim 2.8$ arcminutes. There is additional filtering done to minimize the effect of Sco X-1 on the resulting mosaiced image, due to its brightness, and exclude survey sky images in which the reduced chi squared of the image is sub-optimal. 
	
	To calculate the total flat exposure\footnote{The flat exposure is the summation of exposures from each snapshot included in the calculation of a mosaic image. This exposure is not corrected for with respect to the partial coding fraction of a given pixel with respect to a given part of the sky.} for each mosaic image, we interpolate and add the exposures of each snapshot of interest, $k$, in each pixel, $(i,j)$. 
	\begin{equation}
		E_{i,j}=\sum_k e_{i,j,k}
	\end{equation}
	where $E_{i,j}$ is the mosaic image's flat exposure in each pixel and $e_{i,j,k}$ is the flat exposure map of the $k^{\mathrm{th}}$ snapshot of interest once it has been interpolated on the sky grids. 
	To calculate the partial coding vignetting mask we multiply the flat exposure by the partial coding and then interpolate onto the sky grids and add the values in each pixel. This is given as
	\begin{equation}
		P_{i,j}=\sum_k e_{i,j,k}p_{i,j,k}
	\end{equation}
	where $P_{i,j}$ is the mosaic partial coding exposure map and $p_{i,j,k}$ is the partial coding map for the $k^{\mathrm{th}}$ snapshot of interest once it has been interpolated on the sky grids. 
	
	To calculate the standard deviation in the mosaiced image for energy band $l$, $\sigma_{i,j,l}$, we add the inverse variances and then convert back to normal standard deviation. This is given as:
	\begin{equation}
		\sigma_{i,j, l}=\frac{1}{\sqrt{\sum_k {{t_{i,j,k,l}^{-2}}}}}
	\end{equation}
	where $t_{i,j,k,l}=(v_{i,j,k,l}/C_{i,j,k,l})$ is the energy dependent off-axis corrected standard deviation map once it has been interpolated on the sky grids. Here, $v_{i,j,k,l}$ is the standard deviation map of the snapshot in each energy band and $C_{i,j,k,l}$ is the energy dependent off-axis correction that takes the width of the coded mask into account which affects photon attenuation through the mask for off-axis sources.
	
	To calculate the mosaiced sky image for energy band $l$, $S_{i,j,l}$, we add the sky images weighted by their inverse off-axis corrected variance map for each snapshot. This is given as:
	\begin{equation}
		S{i,j,l}=\frac{\sum_k (s_{i,j,k,l}/C_{i,j,k,l}). t_{i,j,k,l}^{-2} }{\sum_k {{t_{i,j,k,l}^{-2}}}} \label{sky_mosaic}
	\end{equation}
	where $s_{i,j,k,l}$ is the sky image in each energy band as obtained from \texttt{batsurvey} and the quantity $(s_{i,j,k,l}/C_{i,j,k,l})$ is interpolated on the sky grids before being multiplied by $t_{i,j,k,l}$. 
	
	\replaced{With the mosaic images constructed, the sky images can be searched for sources in an input catalog by using the HEASoft \texttt{batcelldetect} task, similar to what is done for the snapshot images. The \texttt{batcelldetect} task will search each sky grid image for sources and calculates a number of parameters related to the source, including the rate of the source for the mosaiced image. Using this information, we can also construct the spectra. We create a PHA file as in the case of a single snapshot (taking note that these rates are still already background subtracted) but we are not able to use the \texttt{batdrmgen} task. The DRM for the spectra that are analyzed from mosaiced images is constructed based on recreating the known spectrum of the. Thus, this type of DRM works best for Crab-like sources.}{Once the mosaic sky images have been constructed with the associated standard deviation maps we can search the images for known sources. The process of searching for sources in the mosaic images is identical to what is done in the processing of individual survey data. We can use the HEASoft \texttt{batcelldetect} task to search the mosaic sky images for known sources that have been passed in through an input catalog or for unknown sources that have been detected at some SNR above user-specified value. The output of \texttt{batcelldetect} is a fits file with the detected sources and their properties including the count rate of the detected source in each of the 8 energy bands. The energy dependent count rate information can be used to construct a PHA file where the count rate in each energy band is already background subtracted. To do spectral fitting with these mosaic spectra, we are unable to use the \texttt{batdrmgen} task which was used to analyze spectra from individual survey datasets and instead have to create a DRM that takes the mosaic systematics into account. The DRM that is used for spectra that are produced from mosaiced images is constructed based on recreating the known spectrum of the Crab Nebula where the photon index $\Gamma=2.15$ and the flux of the Crab Nebula from 14-195 keV is $2.44 \times 10^{-8} \funit$ \citep{tueller_22_mo_survey}.}
	
	Using the mosaiced images, we can construct count rate light curves and spectral/flux evolution of the source at the time binned mosaiced image resolution. In prior survey papers, this time binning was always at the 1 month time scale. 
	
	\section{The \ba Code} \label{ba_code}
	{ 
		In this section we outline the \ba \added{\citep{batanalysis}} python package\footnote{This package is open source and is available on github at: https://github.com/parsotat/BatAnalysis }.
		The \ba{} package allows a user to:
		\begin{itemize}
			\item[1.] Download  BAT survey datasets
			\item[2.] Create Custom Source Catalogs
			\item[3.] Process BAT Survey observations
			\item[4.] Conduct Spectral Fitting
			\item[5.] Determine if a source was detected to some user defined threshold
			\item[6.] Create and Process Mosaiced Images 
		\end{itemize}
		\replaced{The \ba package uses HEASoftpy to accomplish the same processing results while allowing users to batch process survey observations.}{The \ba package is a wrapper for HEASoftpy which allows for identical processing of BAT survey data as what is outlined in Section \ref{survey_orig}. }
		Thus, the files produced by the processing of the survey data are also identical to what is produced by the HEASoft pipeline. In this section we elaborate on each of the prior highlighted capabilities of the software. Documentation is included in the \ba{} github repo which goes into the details of how to utilize the software to accomplish the tasks outlined in this section. Additionally, the codes that were used to produce the results in Section \ref{Sec:discussion} provide examples of how to use the python package.
		
		\subsection{Downloading BAT survey datasets} \label{sec:download}
		
		Prior methods of querying and downloading BAT datasets required using the High Energy Astrophysics Science Archive Research Center (HEASARC) for observations that a user is interested in and then downloading the datasets through their web interface. Alternatively, it is possible to use the Astroquery \citep{astroquery} python package to query HEASARC for the data but the user would still need to download the data using the HEASARC web interface. Now, with the \ba{} code, users can both query HEASARC for observations that match some set of criteria and download the appropriate datasets. 
		
		One set of BAT specific criterion that users should be aware of, that cannot be included in typical queries of HEASARC, is the amount of detectors plane area that is exposed to a given RA/Dec coordinate of interest. We recommend that users use Astroquery to obtain a table of observations that may meet their set of observations requirements and then use the \texttt{swiftbat} python package\footnote{https://github.com/lanl/swiftbat\_python} to calculate the area of the detector plane that is exposed to the astrophysical source of interest\footnote{This exposed area calculation does not take into account the number of active detectors or their position. This is a simple filtering to ensure that sources are relatively close to the BAT boresight to minimize the increased noise at low partial coding fractions.}. If this exposed area is greater than some value (taken to be 1000 cm$^2$ for this paper which corresponds to a partial coding fraction of $\sim 19$\%) then that observation ID will be included in those that ultimately get downloaded\footnote{Depending on the number of snapshots in a single survey observation, the size of an individual survey dataset is typically $\lesssim 100$MB while the directory that holds the processed survey results can be as large as $\sim$ a few GB.}.
		
		\subsection{Creating a Custom Source Catalog}
		The \ba code includes a catalog file that includes \replaced{many}{1278} sources\footnote{A text file with all the sources included in the catalog file can be found on the github repository for easy reference.} If there is a source of interest that the user would like to analyze, and the source is not present in the default catalog file then the user needs to add this source to the catalog that will be used later on in the BatAnalysis pipeline, in section \ref{Sec:obs}. This operation is possible with the \ba{} software and allows for the analysis of previously unknown and uncataloged sources.
		
		\subsection{Processing BAT Survey observations}\label{Sec:obs}
		The steps outlined in \ref{survey_orig} are followed by the \ba code which allows us to conveniently call HEASoftpy's \texttt{batsurvey} and other scripts. As a result, the files that are produced by \ba{} are identical to those produced by HEASoft scripts that are used to analyze BAT survey data. \added{Additionally, users can pass in a dictionary that has key/value pairs that correspond to the HEASoft parameters and the values that the user would like to set for those HEASoft script parameters\footnote{In general, python dictionaries are the means in which users pass parameters to the relevant HEASoftpy tasks within the \ba{} code.}} These properties of the \ba{} code mean that much of the documentation on the HEASoft scripts are still valid for the \ba{} software.
		
		As is outlined in Section \ref{survey_heasoft}, it is also possible to include pattern noise maps\footnote{These noise maps can be obtained from: https://zenodo.org/record/7595904\#.Y9q7pS-B3T8 \citep{noise_pattern_maps}.The pattern maps included in this current release  are only updated until 2019-07-31 however these files will be updated regularly in the future. } when processing survey data, which is necessary if the user plans on making mosaiced images. The code searches for the appropriate pattern noise map to include in the call to \texttt{batsurvey} and if the date of the observation exceeds the most up-to-date pattern noise map, then it uses this pattern noise map by default. This decreases the sensitivity of the mosaic images slightly since the proper daily noise map is not applied however the long term noise trends of the most recent pattern noise maps still get applied removing most of the buildup of pattern noise. If the code cannot find the directory with the pattern noise maps, then it does not incorporate any of these maps in the processing of the survey data. 
		
		Once the survey dataset has been processed, the \ba{} software allows the user to calculate the spectrum (or PHA file) of the source of interest. The \ba package also allows the user to calculate the DRM for each spectrum that has been generated. Other information relevant to sources of interest are also calculated by the \ba{} code including the count rate, local background variance, and the SNR in each of the 8 energy bins used for the survey data and the energy-integrated 14-195 keV energy range.  
		
		\subsection{Spectral Fitting}
		Once a user has created the PHA and DRM files for an given source and observation, \ba{} facilitates a spectral fitting of the spectrum using the pyXspec package. Using pyXspec, \ba will try to fit a model to the spectrum and obtain errors on the various free parameters. The default model is a \texttt{cflux*powerlaw} model although the user can pass in different models and parameters, following the functionality offered by pyXspec. An important aspect of spectral fitting is the type of statistics that is used and \ba also allows the user to employ chi squared or cstat statistics based on the photon count. If the user wants to do their own spectral fitting using other tools they can use the produced DRM and PHA files but may need to take special care with including the appropriate systematic errors. 
		
		\subsection{Detection and non detection criteria}
		The \ba code also allows users to determine if a source was detected to some defined level. If it is not detected or the spectral fit is not well constrained then the code recomputes the spectrum using the background variance times the significance that the user requests. Then, this spectrum is fitted with a powerlaw profile with a fixed photon index specified by the user to get a flux upper limit. This procedure is outlined in \cite{Laha_2022_FRB}. This upper limit estimate is very basic, but more advanced upper limits calculations can easily be undertaken by the user.  
		
		\subsection{Mosaiced Images}
		To create mosaic images, the \ba{} code allows users to create a list of survey observations that they would like to bin into some set of user defined time bins. Then \ba{} will create a mosaic image for each requested time bin assuming that there are survey data to be binned. Additionally, the code also creates a time-integrated mosaic image that extends from the start to the end of the user specified set of time bins. These images are the intermediate summed quantities in Section \ref{mosaic_orig} and the final images with physical units. The \ba pipeline automatically saves the intermediate mosaiced images, which can be summed at a later point in time to produce mosaic images over larger time bins. 
		
		Following the creation of these mosaic images, the \ba{} code allows users to produce the same information as is possible with individual survey observations. 
		
		\subsection{Parallelized Analysis}
		The \ba{} software also has convenient parallelized functions that allow for the expedited analysis of large sets of survey data. \added{These functions allow for multiple survey datasets to be processed at the same time and for multiple mosaic analyses to be conducted at the same time on differing CPUs.} The inputs to these parallelized functions are simplified compared to their full capabilities to allow these functions to operate more generally. However, users can use these functions as templates for their own personal codes which meet their specific analysis needs. 
	}
	

	\section{Applications of BAT Survey data}\label{Sec:discussion}
	In this section, we show checks of the code by analyzing the Crab Pulsar Nebula. Additionally, we show the capabilities of the \ba code as applied to known cataloged sources, such as NGC 2992, and previously unknown, uncataloged sources such as MAXI J0637-430.\footnote{All the code for the examples shown here can be found in the notebooks subdirectory of the github.}
	
	\subsection{Crab Nebula}
	We first look at the Crab Nebula to test the \ba software and verify that it is able to reproduce prior analyses done by \cite{tueller_22_mo_survey}. 
	
	We use astroquery to query the HEASARC for observations of the RA/Dec coordinates of the Crab Nebula from 2004-12-15 to 2006-10-27, the start and end dates of the 22 month survey paper \citep{tueller_22_mo_survey}. We then filter the table of returned survey observations based on observations where the exposed area of the BAT detector plane to the coordinates of the Crab Nebula was at least 1000 cm$^2$. These survey data are then binned and combined appropriately to create monthly and weekly mosaics.
	
\begin{figure}
		\centering
		\includegraphics[width=0.5\textwidth]{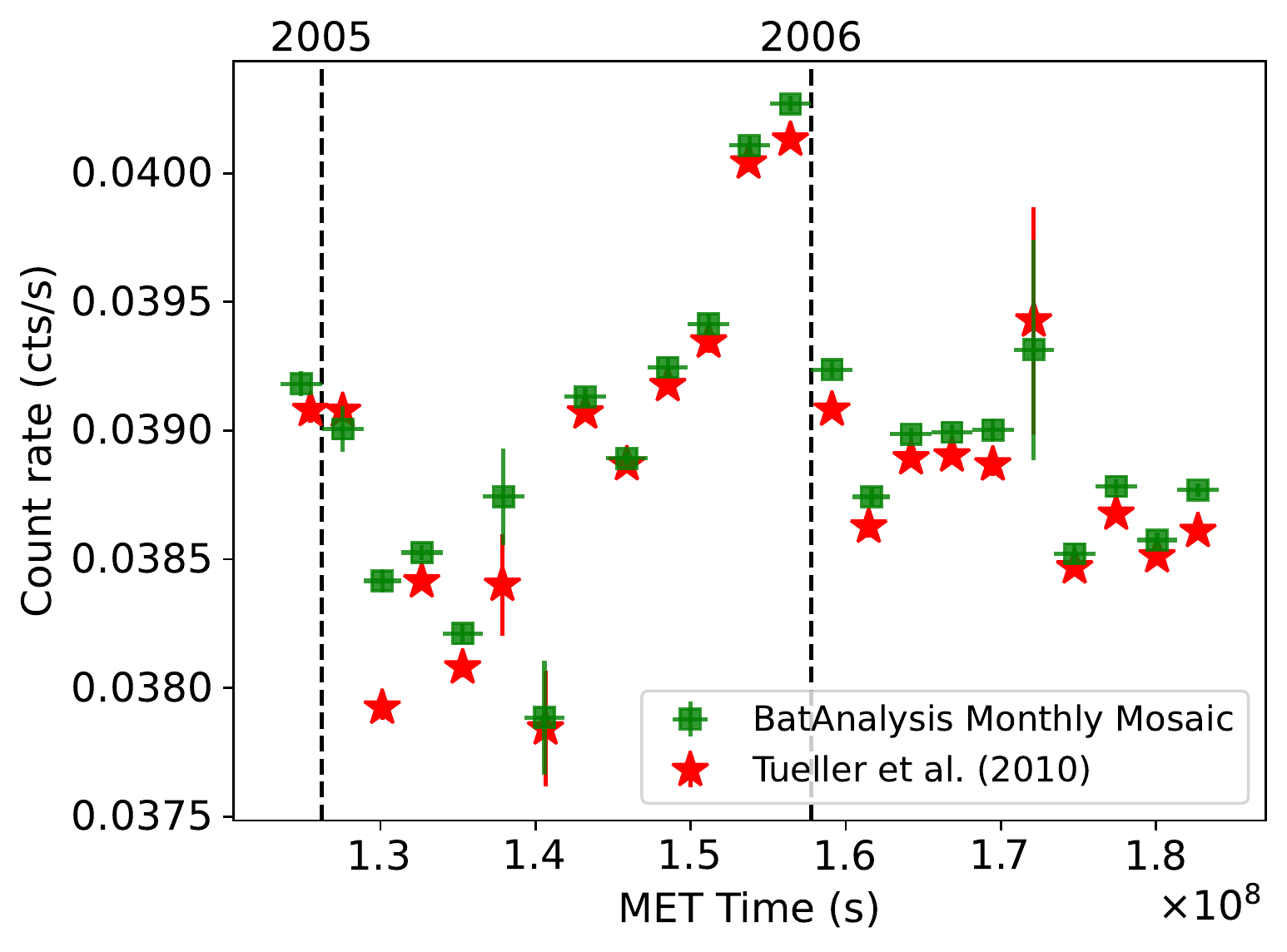}
		\caption{The comparison between the Crab  Nebula monthly mosaic light curve from the 22 month analysis done by \cite{tueller_22_mo_survey}, shown in red, and the rates light curve for the 22 month analysis conducted with the \ba code shown in green. The vertical dashed lines \deleted{in each panel} denotes the start of 2005 and 2006 for reference. The maximum percent difference between the two analyses is 2\% for the month of Feb 2005 showing the ability for the \ba code to recover prior BAT survey analyses.  }
		\label{fig:crab_comparison}
	\end{figure}

	As shown in Figure \ref{fig:crab_comparison}, the monthly mosaic count rate light curve calculated with the \ba code agrees with the results of the 22 month survey \citep{tueller_22_mo_survey} within a few percent, verifying the correctness of the \ba code. We  plot the 22 month survey count rate light curve from \cite{tueller_22_mo_survey} in red while the green points show the monthly mosaiced light curve for the Crab Nebula as calculated with the \ba software. There is a systematic offset of the \ba points with respect to the 22 month survey count rates. This difference can be explained by the underlying data that is used to calculate the mosaic images. In the 22 month survey analysis \cite{tueller_22_mo_survey} use all available survey datasets some of which are noisier than others due to the location of the Crab Nebula being at lower partial coding fractions. Averaging these noisy images with the less noisy images causes the count rate of the Crab Nebula to be lowered since the count rate for sources can become negative once the background is subtracted from noisy images. This effect can be seen in Equation \ref{sky_mosaic}.  With the \ba results, we avoid including noisy images in our analysis by only downloading survey datasets from HEASARC where the Crab Nebula has a partial coding $\ge 19\%$. Due to the lack of noisy images, the measured count rates are slightly higher although this amount is negligible, as can be seen in Figure \ref{fig:crab_comparison}. 
	
\begin{figure*}
		\centering
		\includegraphics[width=\textwidth]{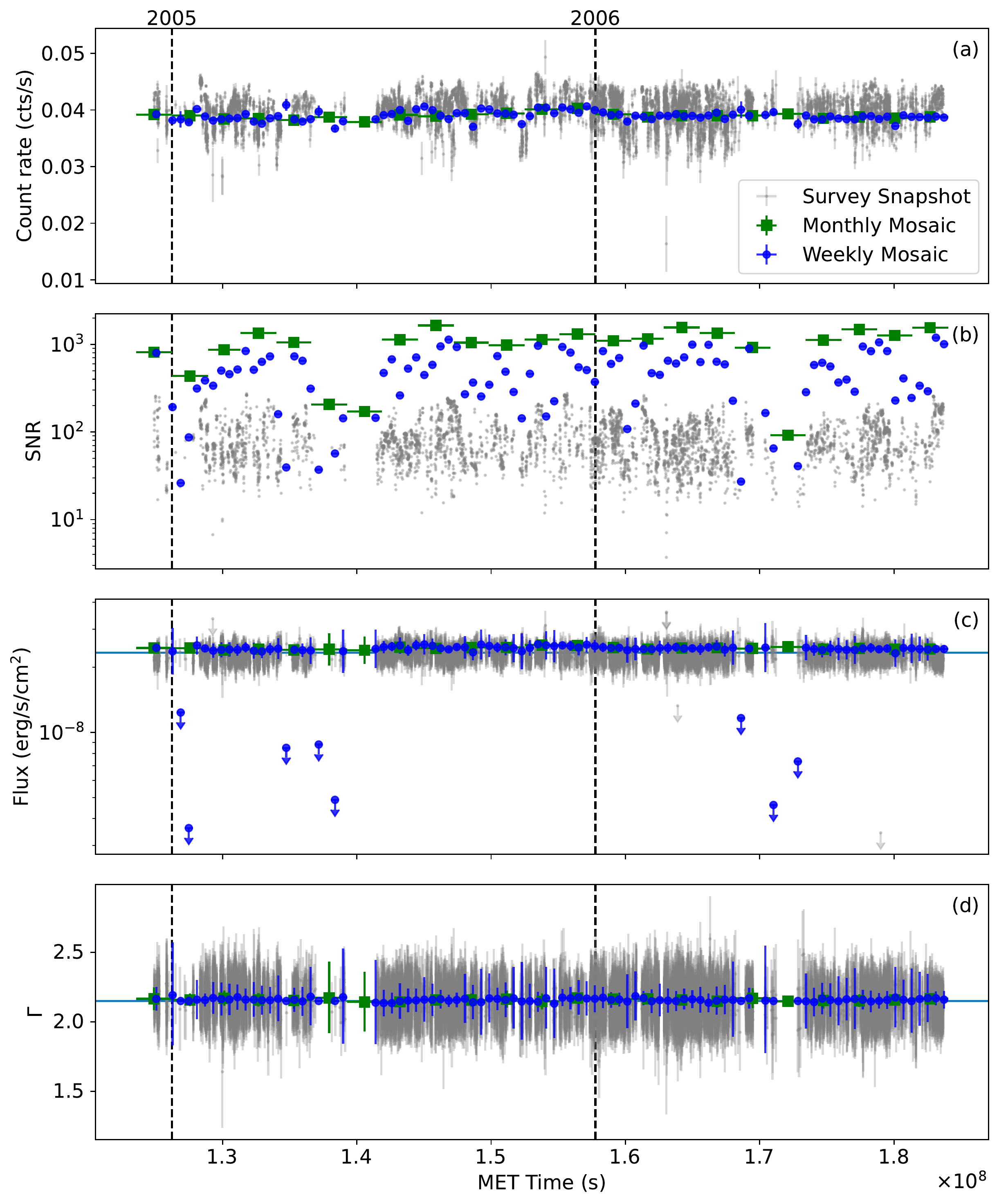}
		\caption{The time varying properties of the Crab Nebula. Panel (a) shows the count rate of the survey snapshots in gray, the weekly mosaic rates, in blue, and the monthly mosaic calculated with the \ba code, shown in green. The survey snapshot data was obtained by querying HEASARC for data spanning the time interval of the BAT 22 month survey \citep{tueller_22_mo_survey}. Panel (b) shows the detected SNR of the Crab Nebula in each binning (snapshot, weekly and monthly). Panel (c) shows the measured flux of the Crab Nebula in each binning as well, where $5\sigma$ upper limits due to poor spectral fits are shown with downward pointing arrows. The horizontal line shows the measured flux of the Crab Nebula, $2.33 \times 10^{-8}$ erg/s/cm$^2$, from the BAT 157 month survey (Lien et. al., in Prep). Panel (d) shows the photon index, $\Gamma$ of the spectral fit, where we freeze $\Gamma=2.15$ in the calculation of the upper limits of the flux when the spectrum is not well fit. The horizontal line shows $\Gamma=2.15$. The vertical dashed lines in each panel denotes the start of 2005 and 2006 for reference.}
		\label{fig:Crab_fig}
	\end{figure*}

	In Figure \ref{fig:Crab_fig} we show the results of analyzing the first 22 months of BAT survey data with the \ba code and the flexibility of the code. In Figure \ref{fig:Crab_fig}(a), we show the count rate light curve of the Crab Nebula at different time resolutions. In gray we show the individual BAT survey snapshots that were downloaded and analyzed, while in green and blue we show the count rates obtained from monthly and weekly mosaiced images. In \ref{fig:Crab_fig}(b) we also show the SNR of these detections of the Crab Nebula at each time resolution. For each survey snapshot and mosaic, we also extracted spectra and fitted them with a \texttt{cflux*po} model and plot the fitted fluxes and photon indexes, $\Gamma$, in Figures \ref{fig:Crab_fig}(c) and \ref{fig:Crab_fig}(d) respectively. Any spectra that were not well constrained were used to estimate $5\sigma$ upper limits on the Crab Nebula at that time period. These points are flux upper limits represented as triangles with $\Gamma=2.15$ which is the value used in the power law spectral fit used to obtain the upper limits. The horizontal line in Figure \ref{fig:Crab_fig}(d) shows $\Gamma=2.15$. 
	
	Additionally, Figure \ref{fig:Crab_fig} shows the survey datasets that are included in various mosaic time bins. Some mosaic time bins have many survey datasets and the detections are strong with well constrained spectra, while others have no or few survey datasets included in their calculation, which affects the resulting SNR and spectral fits. 
	
	\replaced{Scientifically, we}{We} can see the variations that arise on different time scales for the Crab Nebula. The survey snapshots show the greatest variation while the weekly mosaic captures some of this variation with better statistics. Despite the count rate variation, at each timescale, the flux and $\Gamma$ are constant. The spectral photon index for the survey snapshots are centered around the value of $\Gamma=2.15$ however the response files for spectra calculated from mosaiced images are created such that the spectrum of the Crab Nebula reproduces the expected $\Gamma=2.15$, which is also recovered here. The upper limits shown for the Crab Nebula are the result of the Xspec fits not being well constrained, especially since the Crab Nebula is detected at large SNRs. The background variance for the Crab Nebula is so small that the 5$\sigma$ upper limit is well below the accepted flux of the Crab Nebula. These upper limits are shown as example points that a user may obtain in their analyses. They would indicate the need to individually fit these spectra instead of including them in the batch analysis pipeline possible with the \ba package. These individual spectral fittings can be done with the \ba package, but the fitting methods are still relatively simple. 
	
	\subsection{Active Galaxy, NGC 2992}\label{sec:AGN}

	AGN are panchromatic emitters with a significant amount of flux in the hard X-ray band ($>10\kev$). The source of this emission is primarily due to Compton up-scattering of lower energy photons to higher energies by the hot coronal plasma ($T\sim 10^9$K). The resulting spectrum is a power law with a typical slope of $\Gamma\sim 1.5-2.5$ \citep{ricci2017,lahaLLQSO2018,lahaWAXII2016,laha2021,kamraj2022}  extending from the soft to the hard X-rays, with a cut-off energy ($E_{\rm cut}$) in the hard X-rays which corresponds to the coronal plasma temperature ($kT_{\rm e}$) as $E_{\rm cut}\sim 2kT_{\rm e}$, where $k$ is the boltzmann constant. Hence, hard X-ray coverage in the $\sim 10-300\kev$ energy range is crucial to constraining $E_{\rm cut}$ and the coronal plasma temperature. BAT's energy coverage of $14-195\kev$ is ideally suited to detect this spectral component. An additional advantage to observing AGN in the $\sim 10-300\kev$ energy band is that this band is minimally contaminated by absorption due to intervening matter, giving us a direct glimpse of the source.  In this section we demonstrate how the \ba{} pipeline can be used to calculate spectra for an AGN in the crucial energy range $14-195\kev$. 
	
	We use the \ba{} code to calculate the light curve of NGC 2992 using individual survey observations and the monthly mosaic images. NGC 2992 is a well studied bright nearby ($z=0.00771$) AGN, with a typical hard X-ray flux of $F \sim (0.6-12)\times 10^{-11}\funit$ \citep{laha2020,laha2014,middei2022,marinucci2020}. We used survey data from December 2004 to December 2005 and create monthly mosaics for this same time period. We also combine the monthly mosaics into a year long mosaic image and extract a year-averaged spectrum for this AGN.

\begin{figure}
		\centering
		\includegraphics[width=0.5\textwidth]{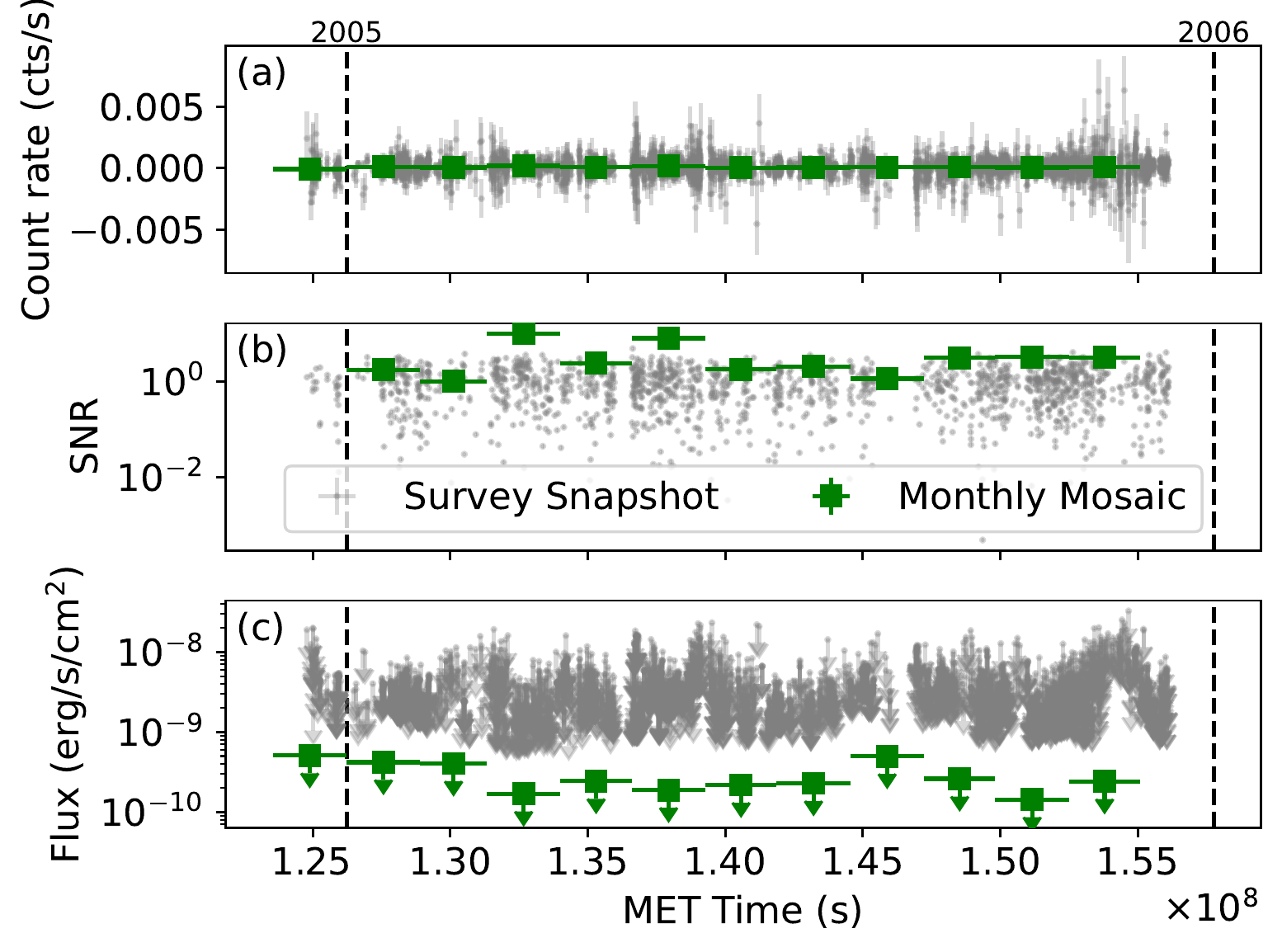}
		\caption{The light curve of the AGN NGC 2992 for individual survey observations (grey dots) and monthly averaged/mosaiced (green dots). Panels (a), (b), (c) are similar to the corresponding panels of Fig \ref{fig:Crab_fig}. The vertical dashed lines in each panel denotes the start of the years 2005 and 2006 for reference. We note that we do not detect the source at a daily or monthly cadence and hence we have upper-limits on the respective quantities. The upper limits were derived using a power law spectrum with a frozen photon index of $\Gamma=1.9$.}
		\label{fig:NGC2992_lc}
	\end{figure}

\begin{figure}
		\centering
		\includegraphics[width=0.5\textwidth]{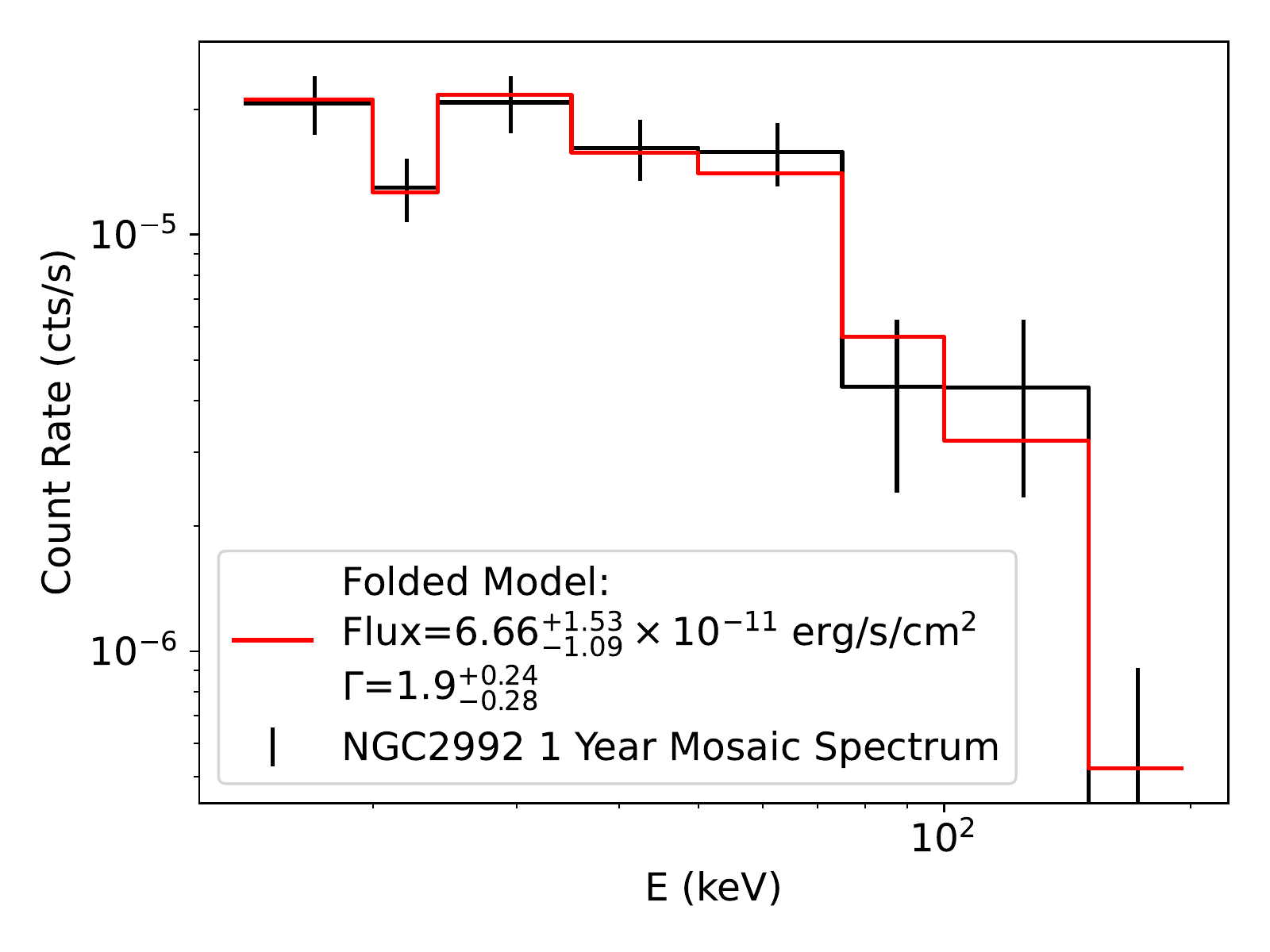}
		\caption{The 1 year averaged (Dec 2004- Dec 2005) BAT spectrum of NGC 2992, calculated using the mosaic technique described in Section \ref{mosaic_orig}. The best fit power law slope and the estimated flux are quoted in the figure, and are consistent with earlier estimates from other hard X-ray telescopes. See Section \ref{sec:AGN} for details.}
		\label{fig:NGC2992_spectra}
	\end{figure}

	The light curve obtained using \ba{} code is shown in Fig \ref{fig:NGC2992_lc}, where the count-rate (top panel), SNR (middle panel) and the $14-195\kev$ flux (lowest panel) are obtained for individual survey observations (grey dots) and also monthly mosaiced images (green squares). Given the low flux threshold of BAT, it is evident that the source has not been detected at a snapshot or monthly cadence and we place $5\sigma$ upper limits on the AGN. With the stacked survey data for a period of 1 year (Dec 2004- Dec 2005) we find that the source is detected and a spectral fit can be constrained (See Fig \ref{fig:NGC2992_spectra}). The flux in the hard band is estimated to be $(6.66\pm 1.5)\times 10^{-11}\funit$, consistent with those measured with other hard-band telescopes such as \nustar{} \citep{middei2022} and Suzaku HXD-PIN \citep{laha2020}. The powerlaw slope measured $\Gamma=1.9\pm 0.25$ is also consistent with previous studies with other hard X-ray telescopes \citep{laha2020}. The $14-195\kev$ spectrum thus obtained (as shown in Fig \ref{fig:NGC2992_spectra}) can be simultaneously fitted with soft X-ray $2-10\kev$ spectrum obtained with other telescopes such as \xmm{} and \chandra{}, thereby carrying out broad band X-ray spectroscopy. The only other operating hard X-ray ($>10\kev$) instrument with imaging capability is \nustar{} (with focusing optics) and that too does not cover energy ranges $>79\kev$. Hence, BAT survey observations provide crucial information in this energy range. 
	
	
	\subsection{Unknown transients}
	Here, we demonstrate the capability for using the \ba code to try to measure detections of an unknown transient in the BAT survey data. We have chosen the source MAXI J0637-430 to analyze as it was a notable new transient discovered by MAXI which was not searched for in BAT survey data. We show what this search looks like in this section and the upper limits we are able to place on this source in the 14-195 keV energy range. 
	
	MAXI J0637-430 was first detected at 06:37 UT on 2019 November 02 \citep{maxi_source} and underwent spectral changes during the month of January 2020 before settling into a hard state by the end of the month \citep{maxi_source_analysis}. This source was identified to be an X-ray Binary with a Black Hole due to timing and spectral analyses \citep{nicer_maxi_source_analysis}. 
	
\begin{figure}
		\centering
		\includegraphics[width=0.5\textwidth]{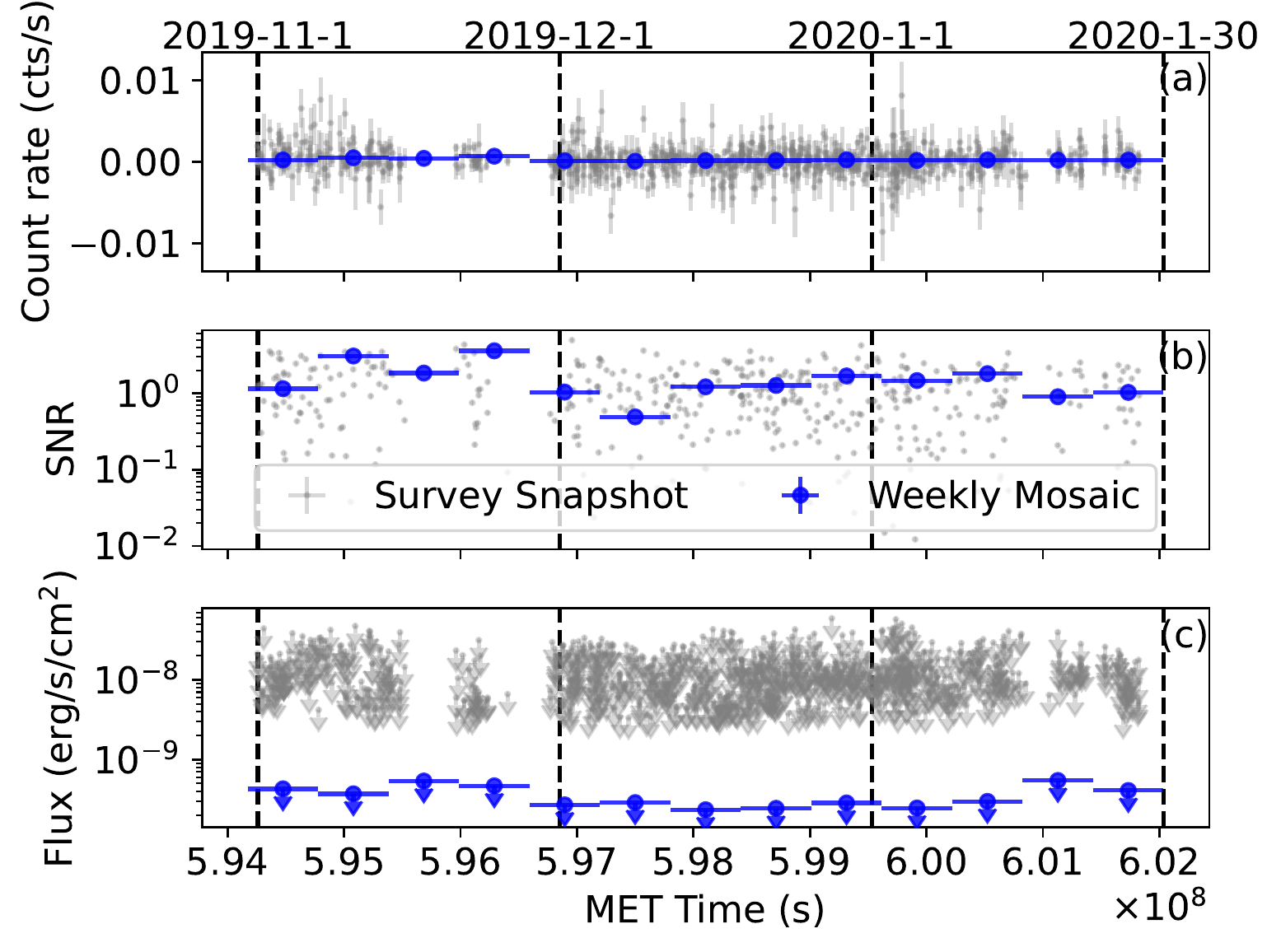}
		\caption{The upper limits that were obtained for MAXI J0637-430 from 14-195 keV from Nov 1st 2019 to Jan 30th 2020. The gray points are the individual survey data sets analyzed and the blue points are the data from analyzing weekly mosaic images. The panel (a) shows the count rate of the source which is consistent with noise. The SNR of the measurements are shown in panel (b). The upper limits derived from the survey and weekly mosaic datasets are shown in panel (c).}
		\label{fig:maxi}
	\end{figure}

	Using the \ba code, we queried HEASARC for survey observations of the coordinates RA/Dec = 99.098, -42.868 from Nov 1st 2019 to Jan 30th 2020 where the point on the sky was exposed to an area of the BAT detector plane of at least 1000 cm$^2$. We processed the survey observations and created weekly mosaics which we analyzed for the MAXI source. Figure \ref{fig:maxi} shows the results of our analyses. The gray points show the analyzed survey data while the blue points show the weekly mosaic analysis. Overall the MAXI source is not significantly detected in the survey or the mosaic analyses with SNR of a few, as shown in Figure \ref{fig:maxi}(b). The \ba code automatically tries to fit the spectra and, upon getting unconstrained spectral parameters, attempts to place 5$\sigma$ upper limits on the flux of the source. Here we freeze the photon index at $\Gamma=2$ to obtain the upper limits. In the survey data, the 5$\sigma$ upper limits are $\sim 10^{-8}$ erg/s/cm$^2$, as show in Figure \ref{fig:maxi}(c). On the other hand, the longer integration times offered by the weekly mosaics allow for deeper limits to be placed at $\sim$ few $\times 10^{-10}$ erg/s/cm$^2$.
	
	\section{Conclusions}\label{Sec:conclusions}
	We have introduced the \ba software, an open source software for processing and analyzing BAT survey data. This dataset has historically been under-utilized however this software allows for a convenient way to download, process, and analyze this dataset unlocking ~18 years worth of archival BAT data. We have used the \ba software to analyze the Crab Nebula and verified that we can recover the 22 month survey results obtained by \cite{tueller_22_mo_survey}. We also exhibit the capability for the code to produce analyses at different time binnings which was not possible before. We also showed examples of using the \ba code to analyze a nearby AGN, NGC 2992, and a previously unknown transient, MAXI J0637-430. We find that the $14-195\kev$ flux values and the spectral slopes match well with previous observations using different hard X-ray telescopes such as \suzaku{}-HXD-PIN and \nustar{}, for both the Crab Nebula and the AGN NGC 2992. 
	
	A limitation of this software is that it does not innately allow for the detection of new sources in the survey data however, it is possible for individuals to conduct their own analyses on the data products created by the \ba code. Additionally, the spectral fitting convenience function included with \ba uses simple methods to fit spectra which can sometimes give unphysical results as in the case of placing $5\sigma$ upper limits on the Crab Nebula although it is detected with a high SNR. However, a user can use the created PHA and DRM files produced with the \ba package to conduct their own rigorous statistical fits.

	Overall, the \ba pipeline builds on a wealth of knowledge that has been built by the BAT team and unlocks this unexplored hard X-ray BAT dataset which contains archival data over a baseline of $\sim 18$ years. This is almost equivalent to the advent of an entirely new instrument for the hard X-ray community, as this data was not fully accessible prior to this software. \added{Typical memory constraints for the \ba{} code running on a single CPU core are $\sim 50$ Mb for running the basic analysis of the BAT survey DPH (the portion of the code that follows Section \ref{survey_orig}) and $\sim 10$ GB for the operation of mosaicing the survey datasets (the portion of the code that follows Section \ref{mosaic_orig}).  The analyses of MAXI J0637-430 took $\sim 2$ days to run in a parallelized analysis from downloading data to producing the mosaic analyses on up to 14 virtual cores on a 2019 MacBook Pro with 8 Intel i-9 cores and 64 GB of memory. }
	
	\added{In the future, we will modify this python package to be able to process and analyze BAT Time Tagged Event data. As a result, the \ba software will become a comprehensive open source pipeline for analyzing some of the most important scientific data products produced by BAT. The community can also contribute to the open source project by creating a fork on github and opening a pull request to incorporate their improvements into the official version of the python package. Additionally, the community can post any issues that they encounter on the github repository for expedited resolution of the problems.  In the future, this package may become an official part of HEASoft, making it an official Swift BAT data analysis pipeline. }

	\acknowledgements
	The material is based upon work supported by NASA under award number 80GSFC21M0002. We thank Hitoshi Negoro for suggesting interesting MAXI sources to analyze. We also thank Brian Irby and Abdu Zoghbi for help with HEASoftpy. \added{We thank the reviewer for suggestions that made this manuscript clearer and more accessible.}\\
	
	\software{Astropy \citep{astropy:2013, astropy:2018,astropy:2022},  Astroquery \citep{astroquery}, \added{BatAnalysis \citep{batanalysis}}
		NumPy \citep{numpy}, Matplotlib \citep{matplotlib}, Scipy \citep{scipy}, HEASoft \citep{HEASoft}, swiftbat\_python, Xspec \citep{xspec}}
	
	\bibliographystyle{aasjournal}
	\bibliography{mybib}
	
	\listofchanges
	
\end{document}